\documentclass[11pt]{article}

\usepackage{epsfig}

\newcommand{\beq}{\begin{eqnarray}}
\newcommand{\eeq}{\end{eqnarray}}
\newcommand{\beqn}{\begin{eqnarray*}}    
\newcommand{\eeqn}{\end{eqnarray*}}
\newcommand{\beqa}{\begin{eqnarray}}
\newcommand{\eeqa}{\end{eqnarray}}
\newcommand{\beqan}{\begin{eqnarray*}}   
\newcommand{\eeqan}{\end{eqnarray*}}

\newcommand{\kom}{\omega}
\newcommand{\gom}{\Omega}
\newcommand{\calm}{{\cal M}}
\newcommand{\merde}{M_{\mbox{\tiny E}}}
\newcommand{\rerde}{R_{\mbox{\tiny E}}}

\newenvironment{bibl}{\begin{list}{}
{\itemindent-7mm \listparindent-7mm
\leftmargin7mm \rightmargin0mm \topsep0mm \partopsep0mm
\setlength{\parsep}{\parskip}}
\item[]}{\end{list}}

\begin{document}

\vspace*{2cm}
\LARGE\bf\baselineskip17pt
SOME ASPECTS ON THE OBSERVATION OF THE\\ GRAVITOMAGNETIC CLOCK EFFECT\\
\\

\normalsize\rm
H. Lichtenegger$^1$, W. Hausleitner$^1$, F. Gronwald$^2$ and B. Mashhoon$^3$\\ \\

$^1${\it Institut f\"ur Weltraumforschung,
         \"Osterreichische Akademie der Wissenschaften,
         A-8042 Graz, Austria},

$^2${\it Institut GET,
         Universit\"at Magdeburg,
         D-39106 Magdeburg, Germany},

$^3${\it Department of Physics and Astronomy,
         University of Missouri, Columbia,
         Missouri 65211, USA}\\ \\

ABSTRACT\\

As a consequence of gravitomagnetism, which is a fundamental weak-field
prediction of general relativity and ubiquitous in gravitational phenomena,
clocks  show a difference in their proper periods when moving along identical
orbits in opposite directions about a spinning mass. This time shift is induced
by the rotation of the source and may be used to verify the existence of the
terrestrial gravitomagnetic field by means of orbiting clocks. A possible mission
scenario is outlined with emphasis given to some of the major difficulties
which inevitably arise in connection with such a venture.\\ \\

INTRODUCTION\\

Among the essential predictions of General Relativity,
gravitomagnetism (and gravitational waves) still lack direct
observational evidence. The only direct reference to date for the existence of gravitomagnetism is based on the orbital
data analysis of Lageos I and II, which suggests that the orbital
precession due to the spinning
Earth corresponds with the Einsteinian prediction within $\sim 20\%$ (Ciufolini {\it et al}., 1998).
The first space experiment directly designed to detect the gravitomagnetic field of the Earth will
be Gravity Probe B, which shall measure the Lense-Thirring precession of a couple of test gyros carried by a spacecraft
along a near-Earth polar orbit.
In the following we describe an alternative space experiment to observe the terrestrial gravitomagnetic field,
based on the temporal structure induced by the rotating Earth, and discuss the influence of the static part of the
terrestrial gravitational field on the outcome of such a measurement. \\

THE CLOCK EFFECT\\

In the weak field and low velocity limit, the field equations of General Relativity reduce to a set of Maxwell-type
equations, thereby giving rise in this approximation to a number of phenomena similar to those known in electrodynamics,
among them the well-known Lense-Thirring precession, which is the gravitational analogue of the precession of a
spinning dipole in a magnetic field. However,
the gravitomagnetic field not only makes a test gyro to precess about
the field lines but also affects the motion
of a test body resulting in a difference in the proper period for co- and counter-revolving particles. Again, this can
be easily understood by comparison with electrodynamics: Consider a test charge moving along a circular orbit
about a central charge. When a weak and homogeneous magnetic field with orientation perpendicular to the orbital plane
is switched on, the test charge experiences an additional Lorentz force, either adding to or opposing the Coulomb
attraction, depending on the direction of the motion of the particle and the magnetic field, respectively.
The particle can, however, still circle along the same path as without the field, but now with a somewhat increased
or decreased velocity; exactly the same is expected to happen in the gravitomagnetic case. The reason for this is
the formal equivalence between the Lorentz force and the Coriolis force, i.e. Larmor's theorem and its
gravitational analogue (Mashhoon, 1993). It follows that among two particles along identical but opposite orbits
about a rotating mass, one will move faster than the other, resulting in a difference of their revolution times.
Numerically, for circular equatorial orbits this time difference
(called gravitomagnetic clock effect) to leading order is found to be
(Cohen and Mashhoon, 1993, Gronwald {\it et al}., 1997)
\begin{equation}
\Delta\tau\simeq 4\pi\frac{J}{Mc^2}\approx2\times 10^{-7}{\rm s},
\end{equation}
where $J$ and $M$ are the angular momentum and mass of the central body, respectively, and the numerical value
results upon inserting the appropriate values for the Earth. In comparison, for Sun and Jupiter, which have the
largest specific angular momenta among all solar system bodies, the clock effect amounts to
$\Delta\tau_{\rm Sun}\simeq 7\times 10^{-5}$ s and $\Delta\tau_{\rm Jupiter}\simeq 5\times 10^{-5}$ s, respectively.\\

Within the PPN formalism, for semiconservative theories and ignoring the Whitehead term and any preferred location effects,
an additional factor of $(1+\gamma+\alpha_1/4)/2$ appears in Eq. (1) (Mashhoon {\it et al}., 2000), showing that the
clock effect depends, as expected, on the same PPN parameter combination as the gravito\-magnetic precession of a gyroscope
(e.g. Will, 1995).\\

It should be emphasized that Eq. (1) does not contain Newton's gravitational constant
$G$; in fact, this is the reason for the large numerical value in (1) with regard to relativistic standards. Another
remarkable characteristic of relation (1) is the independence of the clock effect from the distance of the orbiting
satellites, a feature that is to some extent reminiscent of the Aharonov-Bohm effect. Finally, the positive sign in Eq. (1)
indicates that the corotating satellite is slower than the counterrotating one, apparently in contradiction to the
Machian idea of the dragging of inertial frames. More about these interesting peculiarities can be found in
Mashhoon {\it et al.}, 1999.\\

GRAVITATIONAL PERTURBATIONS\\

In principle, the validity of Eq. (1) can be tested by means of two clocks carried aboard two satellites along identical
but opposite trajectories about the Earth, where each clock is read out when it has exactly covered an azimuthal angle of
$2\pi$.
For an orbit of $7000$ km radius, the satellite travels $\sim 0.04$ mas in 200 ns, whence the accuracy in the
determination of the azimuthal closure must be at least of the same order, which makes extremely high demands on
the tracking facilities. However, assuming no systematic errors in the position measurements, these requirements
become less stringent with the increasing number of revolutions because of the accumulation of the gravitomagnetic
clock effect. Another source of errors is due to the gravitational and non-gravitational
perturbations on the satellites (Lichtenegger {\it et al}., 2000). While the latter may be avoided by means of
drag-free satellites, the former must be carefully modeled in order to extract the gravito\-magnetic effect out of
the clock data. In the following, we will concentrate only on the variation of the orbital period due to the static
part of the Earth's gravitational field (an analysis of the dynamical part will be given elsewhere).\\

Eq. (1) applies to the difference in the sidereal orbital period of two oppositely circling satellites. The angular
velocity of a satellite with respect to a fixed system is the sum
$\dot{\kom}+\dot{\gom}\cos i+\dot{\calm}$  (see Iorio, 2000),
where $\dot{\kom}$ and $\dot{\gom}$ are the time rates of the argument
of perigee and the line of nodes,
respectively, $\dot{\calm}$ is the (perturbed) mean motion and $i$ denotes the orbital inclination.
Upon expansion of the terrestrial gravity field in terms of spherical harmonics, the perturbation equations for the
angular orbital elements read (e.g. Seeber, 1989)
\beqa
\frac{d\kom_{nmpq}}{dt}&=&\frac{GM\rerde^n}{\bar{n}a^{n+3}}
 \left(\frac{\sqrt{1-e^2}}{e}F_{nmp}G'_{npq}-\frac{\cot i}{\sqrt{1-e^2}}F'_{nmp}G_{npq}\right)S_{nmpq}\;,\nonumber\\
\frac{d\gom_{nmpq}}{dt}&=&\frac{GM\rerde^n}{\bar{n}a^{n+3}}\frac{F'_{nmp}G_{npq}S_{nmpq}}{\sqrt{1-e^2}\sin i}\;,\\
\frac{d\calm_{nmpq}}{dt}&=&\bar{n}+\frac{GM\rerde^n}{\bar{n}a^{n+3}}F_{nmp}S_{nmpq}
 \left[2(n+1)G_{npq}-\frac{1-e^2}{e}G'_{npq}\right].\nonumber
\eeqa
Here, $n$ and $m$ are degree and order of the expansion, while $p$ and $q$ are summation indices running from $0$ to
$n$ and $-\infty$ to $\infty$, respectively. Further, $a,\,e$ and $\bar{n}$ denote the semi-major axis, eccentricity
and the unperturbed mean motion of the satellites, and $\merde$ and $\rerde$ are the mass and mean radius of the Earth. Finally,
$S_{nmpq}(\kom,\gom,\calm)$ is a periodic function of the angular coordinates and $F_{nmp}(i)$ and $G_{npq}(e)$ represent
the inclination and eccentricity functions, respectively, where a prime
indicates differentiation with respect to the argument (Kaula, 1966). Secular perturbations are induced by the even zonal
harmonics and affect the variables $\kom,\,\gom$ and $\calm$ while leaving $a,\,e$ and $i$ unchanged, i.e. keeping
the right hand side of (2) constant.
Because $G_{npq}(e)$ represents an infinite series in $e$ with its lowest power equal to $|q|$, for circular orbits
$G_{npq}(e=0)=\delta_{0q}$, where $\delta_{ij}$ is the Kronecker delta. Thus, to first order, the orbital period
modified due to secular perturbations becomes in the limit $e\rightarrow 0$ by means of Eqs. (2)
\beqn
P_{n0p0}\simeq\frac{2\pi}{\bar{n}}\left[1-2(n+1)\frac{\rerde^n}{a^n}F_{n0p}C_{n0}\right].
\eeqn

In the following, in accordance with Eq. (1), we will consider orbits with $e=i=0$ only.
The most relevant perturbation by far is due to $C_{20}$, i.e. the lowest zonal harmonic, resulting in a
difference with respect to the unperturbed period of $\Delta P_{20}^{sec}\simeq -15.7$ s for an orbit of 7000 km
radius.
Because the time dependence of Eqs. (2) is contained in the function $S$, we can estimate an upper limit for
the change in the revolution time due to periodic perturbations by replacing $S$ with its amplitude
$(C_{nm}^2+S_{nm}^2)^{1/2}$ and again calculating the difference from the unperturbed motion.
The first non-vanishing periodic contribution is due to $n=m=2$, having an "amplitude" of
$|\Delta P_{22}^{max}|\simeq 0.158$ s with a period of the perturbation of $P^{Pert}_{22}\simeq 52$ min for a
prograde orbit. In fact, only those combinations of $n$ and $m$ do not vanish for which the sum $n+m$ is even.
Based on NASA's EGM96 gravitational field model (Lemoine \emph{et
al.}, 1998), Table 1 lists all non-vanishing secular contributions to
the orbital
period as well as the amplitudes (in conformity with the above meaning) and periods
of the sectorial and tesseral perturbations up to order and degree 6.

\begin{center}

\vspace{4mm}
\begin{tabular}{|c|c|c|c|c|c|c|c|}\hline
$n,m$&2,0&2,2&3,1&3,3&4,0&4,2&4,4\\\hline
$\Delta P_{nm}^{sec}$(s)&-15.7&&&&-0.024&&\\\hline
$|\Delta P_{nm}^{max}|$(s)&&0.158&0.117&0.117&&0.051&0.032\\\hline
$P_{nm}^{Pert+}$(s)&&3117&6233&2078&&3117&1558\\\hline
$n,m$&5,1&5,3&5,5&6,0&6,2&6,4&6,6\\\hline
$\Delta P_{nm}^{sec}$(s)&&&&-0.008&&&\\\hline
$|\Delta P_{nm}^{max}|$(s)&0.008&0.038&0.071&&0.029&0.040&0.027\\\hline
$P_{nm}^{Pert+}$(s)&6233&2078&1247&&3117&1558&1039\\\hline
\end{tabular}
\end{center}

\vspace{3mm}
Table 1. Secular and periodic changes in the orbital period due to the gravitational field up to degree and order 6
for an orbital radius of 7000 km
(the index $+$ denotes a prograde orbit).

\vspace{4mm}

According to Table 1, the perturbation in the orbital period of a
satellite due to the nonsphericity of the Earth can be up to 8
orders of magnitude larger than the gravitomagnetic
"perturbation". The accurate modeling of the gravitational
perturbations of the Earth is limited by the uncertainty in the
harmonic coefficients $C_{nm}$ and $S_{nm}$. Figure 1a illustrates
the mismodeling $\delta P$ of the period induced by the secular
perturbations up to degree 100, based on the accuracy of the even
zonal harmonic coefficients given in the EGM 96 model. The three
lines correspond to an orbital radius of 7000, 8000 and 12000 km,
respectively. As can be seen, for a near Earth orbit the accuracy
of about the first hundred spherical harmonics is less than
required, while for a 12000 km orbit only the accuracy of the
coefficients up to degree $\sim 20$ are of importance. Figure 1b
shows the maximum mismodeling of the periodic perturbations due to
the uncertainty of the harmonic coefficients $C_{n2}$ and
$S_{n2}$. For a 7000 km orbit, the error is up to 1000 times
larger than the effect being studied and decreases to a factor of
100 at a distance of 12000 km. However, in spite of these large
errors, the influence of the periodic perturbations is of less
significance because they will cancel out as well as will be
masked by the clock effect after a sufficient number of
revolutions due to its accumulative character.

\begin{figure}[t]
 \centerline{\epsfxsize=14cm \epsfbox{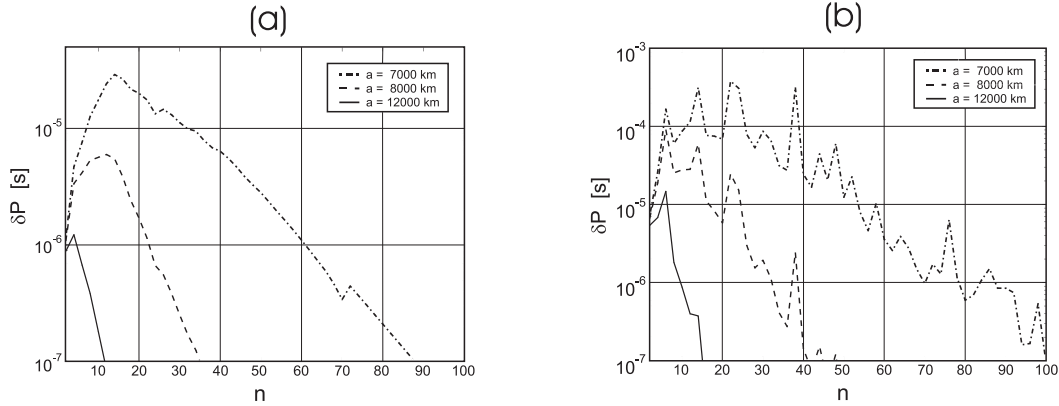}}
 \caption{Mismodeling of the period for three different orbital
 heights: (a) error due to the even zonal perturbations $C_{n0}$,
 (b) error due to the uncertainty in the  coefficients $C_{n2}$ and $S_{n2}$.}\label{fig1}    
\end{figure}

\vspace{7mm}

SUMMARY\\

The gravitomagnetic clock effect, which involves a coupling between the orbital motion of a satellite and the
rotation of the Earth, can be considered as an interesting alternative way to verify the existence of the
gravitomagnetic field predicted by General Relativity. We have examined the influence of the static part of the
gravitational field of the Earth on the period of a satellite and have estimated, based on the EGM96 gravity
field model, the expected mismodeling with respect to the observation of the gravitomagnetic clock effect. While
periodic perturbations, which involve a mismodeling of 100-1000 times larger than required, may cancel out,
secular perturbations must be handled carefully. Assuming an experimental error of $\sim 10\%$, we find the
residual uncertainty exceeding the required accuracy by a factor of 100. Thus, a successful observation of the
clock effect has to await a distinct improvement in the determination
of the gravitational field of the Earth. This may be expected
within a couple of years based on the data of the upcoming GOCE mission.\\

\vspace{7mm}

ACKNOWLEDGMENT\\

We thank Lorenzo Iorio for helpful comments.\\

\vspace{4mm}

REFERENCES\\

\begin{bibl}
Ciufolini, I., E.~Pavlis, F.~Chieppa, E.~Fernandes-Vieira, and
J.~P\'erez-Mercader,
\newblock Test of General Relativity and Measurement of the {L}ense-{T}hirring
  Effect with two Earth Satellites,
\newblock {\em Science}, {\bf 279}, 2100--2103 (1998).

Cohen, J.M. and B.~Mashhoon,
\newblock Standard Clocks, Interferometry, and Gravitomagnetism,
\newblock {\em Phys.\,Lett.\,A}, {\bf 181}, 353--358 (1993).

Gronwald, F., E.~Gruber, H.~Lichtenegger, and R.A. Puntigam,
\newblock Gravity Probe {C{\tiny lock}} - {P}robing the Gravito\-magnetic Field
  of the {E}arth by Means of a Clock Experiment,
\newblock in {\em Fundamental Physics in Space}, pp. 29--37, ESA SP-420 (1997).

Iorio, L.,
\newblock Satellite gravitational orbital perturbations and the gravitomagnetic
 clock effect,
\newblock {\em Int.\,J.\,Mod.\,Phys. D}, {in press (gr-qc/0007014)}, (2000).

Kaula, W.M.,
\newblock Theory of Satellite Geodesy,
\newblock {124 pp.}, {Blaisdell Publishing Company}, Waltham, (1966).

Lemoine, F.\,G., \emph{et al.},
\newblock The Development of the Joint NASA GSFC and the National
Imagery Mapping Agency (NIMA) Geopotential Model EGM96,
\newblock NASA/TP-1998-206861 (1998).

Lichtenegger, H.I.M., F. Gronwald and B. Mashhoon,
\newblock On Detecting the Gravitomagnetic Field of the Earth by Means of
 Orbiting Clocks,
\newblock {\em Adv.\,Space Res.}, {\bf 25}, 1255--1258 (2000).

Mashhoon, B.,
\newblock On the gravitational analogue of Larmor's theorem,
\newblock {\em Phys. Lett. A}, {\bf 173}, 347--354 (1993).

Mashhoon, B., F. Gronwald and D.S. Theiss,
\newblock On measuring gravitomagnetism via spaceborne clocks:
 a gravitomagnetic clock effect,
\newblock {\em Ann.\,Physik}, {\bf 8}, 135--152 (1999).

Mashhoon, B., F. Gronwald and H. Lichtenegger,
\newblock Gravitomagnetism and the Clock Effect,
\newblock in \emph{Testing Relativistic Gravity in Space}, edited by
C.\ L\"ammerzahl, C.\,W.\,F.\ Everitt and F.\,W.\ Hehl, Springer,
Berlin (2001).

Seeber, G.,
\newblock Satellitengeod\"asie,
\newblock {489 pp.}, Walter de Gruyter, Berlin (1989).

Will, C.M.,
\newblock Testing Machian Effects in Laboratory and Space Experiments,
\newblock in {\em Mach's Principle: From Newton's Bucket to Quantum Gravity},
 volume~6 of {\em Einstein Studies}, edited by J.~Barbour and H.~Pfister,
 pp. 365--385, Birkh\"auser, Boston (1995).

\end{bibl}

\end{document}